\documentclass[seceq,supplement]{ptptex}

\usepackage{graphicx, color}
\usepackage{latexsym}
\newcommand{\vect}[1]{\mbox{\boldmath${#1} $}} 
\newcommand{\vex}{{\vect x}}

\title{
Transport coefficients of the second order hydrodynamics \\
and the applicability of hydrodynamic model%
}

\author{
Shin \textsc{Muroya}%
}

\inst{
Dept. of C.M. 
Matsumoto Univ. 
Matsumoto 390-1295, Japan
}

\abst{
Based on the Nakajima-Zubarev type nonequilibrium density operator, we derive microscopic formulae of the transport coefficients in the second order hydrodynamics. 
}

\begin{document}

\maketitle

\section{Introduction}

Hydrodynamical model 
is one of the widely applied phenomenological models for the relativistic heavy ion collisions.  
According to the recent detailed analyses, weak but non-vanishing viscosity is significant for
the qualitative discussion of $v_2$ at RHIC.\cite{Hirano:2008hy} 
Because Navier-Stokes equation is a parabolic type equation, naive relativistic 
extension is not consistent with causality and numerical solution becomes unstable.\cite{HL} 
The relativistic causal hydrodynamics had been introduced phenomenologically by Israel and Stewart more than thirty years ago.\cite{IS1, IS2}  In these decades, many proposal appear for the appropriate equations of the relativistic causal hydrodynamics.\cite{TK, KK, MH} 
Most of the works seem to belong semi-classical phenomenological approach based on Boltzmann equation.  In this paper we discuss a relativistic causal hydrodynamics as the second order hydrodynamics  based on a nonequilibrium density operator method.\cite{KYN}

Hydrodynamical model is a phenomenological model of the macroscopic point of view. 
By coarse graining, system is constructed as a patchwork of corresponding 
statistical systems in equilibrium.  
The system at the space-time point is microscopically 
large but macroscopically one point and moving with four velocity $U^{\mu}$.  
Thermodynamical quantities and coefficients of hydrodynamical equation 
should be the local quantities as the 
function of position through the local thermodynamical parameters such as 
temperature $T(x)$ and chemical  potential $\mu(x)$.  
In principle, we can calculate them  by using statistical mechanics.  The aim of this paper is to describe all  coefficients in the second order hydrodynamics as expectation values of the local equilibrium state.  
         
\section{Notations}

Hydrodynamical equation is composed of energy-momentum conservation low, $\partial^{\rho} T_{\rho \sigma} = 0,$ and charge conservation low, $\partial^{\rho} J_{\rho} = 0,$.
Flow of the fluid is described by a four velocity $U^{\mu}$ which is a normalized time-like vector, 
$U^{\mu}U_{\mu} = 1$, and by using $U^{\mu}$ space-like projection operator is defined as, 
$\Delta^{\mu \nu} = g^{\mu \nu}- U^{\mu}U^{\nu}$.

We adopt Landau-Lifshitz frame (E-frame); $U^{\mu}$ is eigenvector of the energy momentum tensor,\cite{LL} 
$
T^{\mu \nu }U_{\nu}. =  \varepsilon  U^{\mu}. 
$
Pressure $P$ is defined by the trace of spatial part of the energy-momentum tensor,
$
P =  \frac{-1}{3}{\Delta_{\mu}}_{\nu}T^{\mu \nu }.
$
Stress-shear tensor is given as the spatial traceless part,
$$
\pi^{\mu \nu} = ({\Delta^{\mu}} _{\rho} {\Delta^{\nu}} _{\sigma} - \frac{1}{3} {\Delta^{\mu}} ^{\nu}{\Delta_{\rho}} _{\sigma})  T^{\rho \sigma}.
$$

Charge density is given by $
n=  J^{\mu}U_{\mu}$ as usual, however,  
 $U^{\mu} $ is proportional to the energy flow on the E-frame, 
charge current contains $I^{\mu}$ which is perpendicular to $U^{\mu} $,
$$
J^{\mu} =  n U^{\mu} + I^{\mu}.
$$
$\tau$ stands for time on the comoving flame of the fluid element, $\tau =  x^{\mu}U_{\mu}$, and 
$D = U^{\mu} \partial_{\mu}$ stands for the derivative by $\tau$, 
respectively.

\section{Nonequilibrium density operator}

Let us focus our discussion to the only small deviation from thermal equilibrium state.
Our aim is to derive equations which enable us to discuss second order deviation
from the thermal equilibrium state.
 
Nonequilibrium density operator is given as,\cite{KYN, CH}
\begin{eqnarray}
\hat{\rho} =  Q^{-1} {\rm exp} \Bigg[  & &\left. \displaystyle{\int}_{ }^{ } dx^{\mu}
\left\{
\beta(\vex,t)
 U^{\nu}(\vex,t)\hat{T}_{\mu \nu}(\vex,t) \right.  
\left. -\beta(\vex,t) \mu(\vex,t) 
\hat{J}_{\mu}(\vex,t) \right\} \right. \nonumber \\
&-&  \lim _{\zeta \to 0+}\displaystyle{\int} d^{4} {\vex'}
{\rm e}^{-\zeta U^{\mu}(x-x')_{\mu}}
\left\{ \hat{T}_{\mu \nu}(\vex,'t')\partial'^{\mu}(\beta(\vex',t') U^{\nu}(\vex',t')) \right.  \nonumber \\
 &&      
\left.+ \hat{J}_{\nu}(\vex',t')\partial'^{\nu}(\beta(\vex',t') \mu(\vex',t')) \right\} \Bigg], \quad
U^{\mu}(x-x')_{\mu} \ge 0 .
\end{eqnarray}
The first term of the right hand side corresponds to the local equilibrium density operator,
$\hat{\rho}_l=Q^{-1}_{l} {\rm exp} ({\int} dx^{\mu}
\{
\beta(\vex,t) U^{\nu}(\vex,t)\hat{T}_{\mu \nu}(\vex,t)-
\beta(\vex,t) \mu(\vex,t) \hat{J}_{\mu}(\vex,t) 
\} ), $
with $ Q_{l} $ being normalization. $dx^{\mu}$ is a space like hypersurface which locally corresponds to a comoving flame of the fluid element. Though out this paper, the limit, $\zeta \to 0+$,  should be taken at the final stage of the calculation.
The second term stands for the effect of thermodynamical force. 
Expectation value of any operator $\hat{O}(x)$ is given by $\langle \hat{O}(x) \rangle = \rm{tr}\{ \hat{\rho} \hat{O}(x) \}$. 

Following the standard procedure of linear response theory, we 
 expand $\hat{\rho}$ around local equilibrium operator, $\hat{\rho}_l$, 
up to the linear term of thermodynamical force.  
Expectation value, $\langle \hat{O}(x) \rangle$,  is given as a local equilibrium expectation 
of canonical commutation with the force term,
\begin{eqnarray}
\langle \hat{O}(x) \rangle =\langle \hat{O}(x) \rangle_{l}  
&+& \displaystyle{
\lim _{\zeta \to 0+}\displaystyle{\int} d^{4} {\vex'}
{\rm e}^{-\zeta U^{\mu}(x-x')_{\mu}} }
\left(\hat{O}(x),\hat{T}_{\rho \sigma}(x'){\partial'} ^{\rho}\left(\beta(x')U^{\sigma}(x')\right) \right) \nonumber \\
&+& \displaystyle{
\lim _{\zeta \to 0+}\displaystyle{\int} d^{4} {\vex'}
{\rm e}^{-\zeta U^{\mu}(x-x')_{\mu}} }
\left(\hat{O}(x),\hat{J}_{\rho }(x'){\partial' }^{\rho}\left(\beta(x')\mu(x')\right) \right), 
\end{eqnarray}
with $(O,B)$ being canonical commutation relation, 
$$
(O,B)=\langle O \displaystyle{\int_0 ^{1}}d\lambda e^{\lambda A} B e^{-\lambda A} \rangle_{l}
- \langle O \rangle_{l} \langle B \rangle_{l}.
$$

\section{Derivative expansion with thermodynamical parameters}

The change of thermodynamical parameters are slow enough and almost constant during the
microscopic relaxation length which is determined by the canonical correlation, we may
adopt Taylor expansion of the thermodynamical force, 
${\partial'}^{\rho}\left(\beta(x')U^{\sigma}(x') \right)$  and 
${\partial'}^{\rho}\left(\beta(x')\mu(x')\right)$
at around position $x$ of the operator $\hat{O}(x)$.  
On the local comoving flame of the fluid element at $x$, we can replace the local equilibrium distribution with the corresponding distribution in equilibrium.  

Then we can obtain expectation values of $T^{\mu\nu}$ and $J^{\mu}$ in the power series of 
gradient of thermodynamical parameters,
\begin{eqnarray}
\langle \hat{T}^{\mu \nu}(x) \rangle &=&\langle \hat{T}^{\mu \nu}(x) \rangle_{l} 
 + \int d^3 \vex ' {\int^{t}}_{-\infty} dt' {\rm e} ^{-\zeta(t-t')}  (\hat{T}^{\mu \nu}(x),
\hat{T}_{\rho \sigma}(x')) \partial ^{\rho} \left(\beta U^{\sigma}\right)  \nonumber \\
 &&+ \int d^3 \vex ' {\int^{t}}_{-\infty} dt' {\rm e} ^{-\zeta(t-t')}  (\hat{T}^{\mu \nu}(x),
\hat{J}_{\rho}(x'))
\partial ^{\rho} \left(\beta \mu\right) \nonumber \\
&&+\int d^3 \vex ' {\int^{t}}_{-\infty} dt' {\rm e} ^{-\zeta(t-t')} (\hat{T}^{\mu \nu}(x),({x_{\lambda}}'-x_{\lambda})
\hat{T}_{\rho \sigma}(x'))
\partial ^{\lambda}
\partial ^{\rho}\left(\beta U^{\sigma}\right) \nonumber \\
&&+\int d^3 \vex ' {\int^{t}}_{-\infty} dt' {\rm e} ^{-\zeta(t-t')} (\hat{T}^{\mu \nu}(x),({x_{\lambda}}'-x_{\lambda})
\hat{J}_{\rho}(x'))\partial ^{\lambda}
\partial ^{\rho}\left(\beta \mu \right), \label{tmn}
\end{eqnarray}
\begin{eqnarray}
\langle \hat{J}^{\mu}(x) \rangle &=& \langle \hat{J}^{\mu}(x) \rangle_{l}  
+ \int d^3 \vex ' {\int^{t}}_{-\infty} dt' {\rm e} ^{-\zeta(t-t')}  (\hat{J}^{\mu}(x),
\hat{J}_{\rho}(x'))
\partial ^{\rho} \left(\beta \mu\right)  \nonumber \\
&& + \int d^3 \vex ' {\int^{t}}_{-\infty} dt' {\rm e} ^{-\zeta(t-t')}  (\hat{J}^{\mu}(x),
\hat{T}_{\rho \sigma}(x'))
\partial ^{\rho} \left(\beta U^{\sigma}\right) \nonumber \\
&&+\int d^3 \vex ' {\int^{t}}_{-\infty} dt' {\rm e} ^{-\zeta(t-t')} (\hat{J}^{\mu}(x),({x_{\lambda}}'-x_{\lambda})
\hat{J}_{\rho}(x'))
\partial ^{\lambda}
\partial ^{\rho}\left(\beta \mu \right) \nonumber \\
&&+\int d^3 \vex ' {\int^{t}}_{-\infty} dt' {\rm e} ^{-\zeta(t-t')} (\hat{J}^{\mu}(x),({x_{\lambda}}'-x_{\lambda})
\hat{T}_{\rho \sigma}(x')) 
\partial ^{\lambda}
\partial ^{\rho}\left(\beta U^{\sigma}\right). \label{jm}
\end{eqnarray}
The first terms of the right hand side of (\ref{tmn}) and (\ref{jm}) are the zeroth order terms 
which correspond to the perfect fluid part, $
\langle \hat{T}^{\mu \nu}(x) \rangle_{l} = (\varepsilon +P)U^{\mu}U^{\nu} - Pg^{\mu \nu}$, 
and
 $\langle \hat{J}^{\mu}(x) \rangle_{l} = n J^{\mu}.$
The second terms of the right hand side provide us Navier-Stokes equation.  Well known Kubo-formula for 
 viscosity and heat conductivity are obtained as the coefficients of the first order gradient of thermodynamical parameters. \cite{KYN, NI} 
 
In both equations, the third terms vanish by virtue of Curie's theorem, 
because of isotropic property of the local equilibrium state.
The forth and the fifth terms stand for the second order deviations.
%
For example, the forth term in right hand side of (\ref {jm}) can be rewritten as, 
\begin{eqnarray}
 &&\int d^3 \vex ' {\int^{t}}_{-\infty} dt' {\rm e} ^{-\zeta(t-t')} 
\left( \hat{J}^{\mu}(x),({x_{\lambda}}'-x_{\lambda}) \hat{J}_{\rho}(x')\right)
\partial ^{\lambda}\partial ^{\rho}\left(\beta \mu \right) \nonumber \\ 
 \noalign{\vskip 0.4cm}
&=&\int d^3 \vex ' {\int^{t}}_{-\infty} dt' {\rm e} ^{-\zeta(t-t')} (\hat{J}^{\mu}(x),(t'-t)\hat{J}_{\rho}(x'))
D \partial ^{\rho}\left(\beta \mu \right)  \nonumber \\ 
&+&\int d^3 \vex ' {\int^{t}}_{-\infty} dt' {\rm e} ^{-\zeta(t-t')} (\hat{J}^{\mu}(x),\Delta^{\sigma \lambda}({x_{\lambda}}'-x_{\lambda})\hat{J}_{\rho}(x'))
\Delta_{\sigma \rho} \partial ^{\lambda}\partial ^{\rho}\left(\beta \mu \right). \label{b1}
\end{eqnarray}
Comparing (\ref{jm}) and (2.38b) in Ref.\ 4, 
the first term of the right hand side of (\ref{b1}) corresponds to the time-derivative term of heat current and the coefficient corresponds to $ \kappa^{2} T^{3} n^{2} {\beta}_{1}/(\epsilon+P)^2$ in Israel-Stewart equation\cite{IS2}. 
On the other hand, 
the second term in (\ref{b1}) must vanish by virtue of Curie's theorem. 

\section{Concluding Remarks}

All coefficients in the derivative expansion in (\ref{tmn}) and (\ref{jm}) are expressed by the canonical correlations of $T^{\mu\nu}$, $ J^{\mu}$ and $x'^{\mu}-x^{\mu}$.  Coefficients for the second order terms are current-weighted relaxation time,
$$
\int d^3 \vex ' {\int^{t}}_{-\infty} dt' {\rm e} ^{-\zeta(t-t')} (\hat{J}^{\mu}(x),(t'-t)\hat{J}_{\rho}(x')),
$$
and  current-weighted correlation length,
$$
\int d^3 \vex ' {\int^{t}}_{-\infty} dt' {\rm e} ^{-\zeta(t-t')}  (\hat{T}^{\mu\nu}(x), \Delta^{\sigma \lambda}({x_{\lambda}}'-x_{\lambda})\hat{J}_{\rho}(x')).
$$
These quantities can be evaluated by using hadro-molecular simulation\cite{MS,Xian }or Lattice field theory.
Comparing these quantities to the scale of the change of thermodynamical parameters
in the hydrodynamical solution, we can investigate the self-consistent check of the 
hydrodynamical picture\cite{IMN}.

\section*{Acknowledgments}
The author would like to thank Dr.\ Masashi Mizutani, Prof.\ Tetsufumi Hirano and Prof.\ Naomichi Suzuki for fruitful discussion.  This work is supported by Grants-in-Aid for Research Activity of Matsumoto University No.\ 11111048.

%

\end{document}